\documentclass[12pt]{article}
\usepackage{fancyhdr}

\usepackage{mathrsfs}
\usepackage[T1]{fontenc}
\usepackage{mathpazo}
\usepackage{setspace}
\usepackage{amsfonts}
\usepackage{amssymb}
\usepackage{amsmath}
\usepackage{epsfig}
\usepackage{latexsym}
\usepackage{color}
\usepackage{graphicx}
\usepackage{nicefrac}
\usepackage[latin1]{inputenc}
\usepackage{slashed}
\usepackage{multirow}
\usepackage{comment}
\usepackage{soul}
\usepackage{hyperref}
\usepackage[nosort]{cite}
\usepackage{datetime}
\usepackage{braket}
\usepackage{float}
\usepackage[normalem]{ulem}
\usepackage[super]{nth}

\usepackage[splitrule]{footmisc}
\usepackage[titletoc]{appendix}

\def\hybrid{\topmargin -20pt    \oddsidemargin 0pt
        \headheight 0pt \headsep 0pt
        \textwidth 6.25in       
        \textheight 9.25in       
        \marginparwidth .875in
        \parskip 5pt plus 1pt   \jot = 1.5ex}

\hybrid

\def\baselinestretch{1.2}

\catcode`\@=11

\def\marginnote#1{}
\newcount\hour
\newcount\minute
\newtoks\amorpm
\hour=\time\divide\hour by60
\minute=\time{\multiply\hour by60 \global\advance\minute by-\hour}
\edef\standardtime{{\ifnum\hour<12 \global\amorpm={am}%
        \else\global\amorpm={pm}\advance\hour by-12 \fi
        \ifnum\hour=0 \hour=12 \fi
        \number\hour:\ifnum\minute<10 0\fi\number\minute\the\amorpm}}
\edef\militarytime{\number\hour:\ifnum\minute<10 0\fi\number\minute}

\def\draftlabel#1{{\@bsphack\if@filesw {\let\thepage\relax
   \xdef\@gtempa{\write\@auxout{\string
      \newlabel{#1}{{\@currentlabel}{\thepage}}}}}\@gtempa
   \if@nobreak \ifvmode\nobreak\fi\fi\fi\@esphack}
        \gdef\@eqnlabel{#1}}
\def\@eqnlabel{}
\def\@vacuum{}
\def\draftmarginnote#1{\marginpar{\raggedright\scriptsize\tt#1}}

\def\draft{\oddsidemargin -.5truein
        \def\@oddfoot{\sl preliminary draft \hfil
        \rm\thepage\hfil\sl\today\quad\militarytime}
        \let\@evenfoot\@oddfoot \overfullrule 3pt
        \let\label=\draftlabel
        \let\marginnote=\draftmarginnote
   \def\@eqnnum{(\theequation)\rlap{\kern\marginparsep\tt\@eqnlabel}%
\global\let\@eqnlabel\@vacuum}  }

\def\preprint{\twocolumn\sloppy\flushbottom\parindent 2em
        \leftmargini 2em\leftmarginv .5em\leftmarginvi .5em
        \oddsidemargin -.5in    \evensidemargin -.5in
        \columnsep .4in \footheight 0pt
        \textwidth 10.in        \topmargin  -.4in
        \headheight 12pt \topskip .4in
        \textheight 6.9in \footskip 0pt
        \def\@oddhead{\thepage\hfil\addtocounter{page}{1}\thepage}
        \let\@evenhead\@oddhead \def\@oddfoot{} \def\@evenfoot{} }

\def\numberbysection{\@addtoreset{equation}{section}
        \def\theequation{\thesection.\arabic{equation}}}

\def\underline#1{\relax\ifmmode\@@underline#1\else
        $\@@underline{\hbox{#1}}$\relax\fi}

\def\titlepage{\@restonecolfalse\if@twocolumn\@restonecoltrue\onecolumn
     \else \newpage \fi \thispagestyle{empty}\c@page\z@
        \def\thefootnote{\fnsymbol{footnote}} }

\def\endtitlepage{\if@restonecol\twocolumn \else \newpage \fi
        \def\thefootnote{\arabic{footnote}}
        \setcounter{footnote}{0}}  

\catcode`@=12
\relax

\usepackage{caption}
\def\figcap{\section*{Figure Captions\markboth
        {FIGURECAPTIONS}{FIGURECAPTIONS}}\list
        {Figure \arabic{enumi}:\hfill}{\settowidth\labelwidth{Figure
999:}
        \leftmargin\labelwidth
        \advance\leftmargin\labelsep\usecounter{enumi}}}
 \relax
\def\tablecap{\section*{Table Captions\markboth
        {TABLECAPTIONS}{TABLECAPTIONS}}\list
        {Table \arabic{enumi}:\hfill}{\settowidth\labelwidth{Table
999:}
        \leftmargin\labelwidth
        \advance\leftmargin\labelsep\usecounter{enumi}}}
 \relax
\def\reflist{\section*{References\markboth
        {REFLIST}{REFLIST}}\list
        {[\arabic{enumi}]\hfill}{\settowidth\labelwidth{[999]}
        \leftmargin\labelwidth
        \advance\leftmargin\labelsep\usecounter{enumi}}}
 \relax
\makeatletter
\newcounter{pubctr}
\def\publist{\@ifnextchar[{\@publist}{\@@publist}}
\def\@publist[#1]{\list
        {[\arabic{pubctr}]\hfill}{\settowidth\labelwidth{[999]}
        \leftmargin\labelwidth
        \advance\leftmargin\labelsep
        \@nmbrlisttrue\def\@listctr{pubctr}
        \setcounter{pubctr}{#1}\addtocounter{pubctr}{-1}}}
\def\@@publist{\list
        {[\arabic{pubctr}]\hfill}{\settowidth\labelwidth{[999]}
        \leftmargin\labelwidth
        \advance\leftmargin\labelsep
        \@nmbrlisttrue\def\@listctr{pubctr}}}
 \relax
\makeatother
\newskip\humongous \humongous=0pt plus 1000pt minus 1000pt

\newif\ifdtup

\relax

\def\be{\begin{equation}}
\def\ee{\end{equation}}
\def\ba{\begin{eqnarray}}
\def\ea{\end{eqnarray}}

\def\del{\partial}

\def\k{\kappa}

\def\G{\Gamma}
\def\d{\delta}
\def\D{\Delta}
\def\e{\epsilon}

\def\th{\theta}
\def\Th{\Theta}
\def\m{\mu}
\def\n{\nu}
\def\om{\omega}

\def\l{\lambda}
\def\L{\Lambda}
\def\s{\sigma}

\def\no{\noindent}

\def\qq{\qquad}

\def\IR{\relax{\rm I\kern-.18em R}}

\def \ha {{1\over 2}}

\def \ov {\over}

\def\diag{{\rm diag}}

\def\IR{\relax{\rm I\kern-.18em R}}
\def\IL{\relax{\rm I\kern-.18em L}}

\def\inv{^{\raise.15ex\hbox{${\scriptscriptstyle -}$}\kern-.05em 1}}

\def\Tr{{\rm Tr}}

\begin{document}

\emergencystretch 3em

\renewcommand{\theequation}{\thesection.\arabic{equation}}
\csname @addtoreset\endcsname{equation}{section}

\newcommand{\beq}{\begin{equation}}
\newcommand{\eeq}[1]{\label{#1}\end{equation}}
\newcommand{\ber}{\begin{equation}}
\newcommand{\eer}[1]{\label{#1}\end{equation}}
\newcommand{\eqn}[1]{(\ref{#1})}
\begin{titlepage}
\begin{center}


{\large\bf Integrable $\lambda$-deformations of the Euclidean black string}

\vskip 0.4in

{\bf Sibylle Driezen}$^a$\ and {\bf Konstantinos Sfetsos}$^{b}$
\vskip 0.17in

{\em${}^a$Instituto Galego de F\'isica de Altas Enerx\'ias (IGFAE),\\ Universidade de Santiago de Compostela, Spain
}

\vskip 0.1in

{\em${}^b$Department of Nuclear and Particle Physics,\\ Faculty of Physics, National and Kapodistrian University of Athens,\\15784 Athens, Greece}

\vskip 0.1in

{\footnotesize \texttt sibylle.driezen@usc.es, ksfetsos@phys.uoa.gr}

Tuesday \nth{26} January, 2021

\vskip .5in
\end{center}

\centerline{\bf Abstract}


\no
Non-trivial outer algebra automorphisms may be utilized in $\l$-deformations of (gauged) WZW models
thus providing an efficient way to construct new integrable models. 
We provide two such integrable deformations of the exact coset CFT $SU(2)_k\times U(1)/U(1)_q$
with a vector and axial residual gauge.
Besides the integer level $k$ and the deformation parameter $\l$, these models are characterized by the embedding
parameter $q$ of the $U(1)$ factor.  We show that an axial-vector T-duality persists along the deformations  and, therefore,  
 the models are canonically equivalent.
We demonstrate integrability even though the space is non-symmetric and compute the $RG$-flow equations for the parameters $\l$ and $q$.
Our example provides an integrable deformation of the gravitational solution representing 
a Euclidean three-dimensional black string.

\no

\vskip .4in
\noindent
\end{titlepage}
\vfill
\eject

\newpage

\tableofcontents

\noindent

\def\baselinestretch{1.2}
\baselineskip 20 pt
\noindent


\setcounter{equation}{0}
\renewcommand{\theequation}{\thesection.\arabic{equation}}


\setstretch{1.2}

\section{Introduction}

After the major success of using worldsheet integrability  to provide rigorous tests for the AdS/CFT correspondence, deformations of two-dimensional $\sigma$-models which preserve integrability has attracted considerable attention in order to describe more general theories. In particular, an interesting application is to reduce the number of (super)symmetries, either in the gauge theories or in the black brane backgrounds, whilst preserving the computational power of integrability.

In this paper we focus on integrable deformations of the $\lambda$-type which  finitely deform two-dimensional $\sigma$-models corresponding to exact conformal field theories (CFTs).  In its original formulation \cite{Sfetsos:2013wia}  the deformation acts on the  Wess-Zumino-Witten (WZW) model on a Lie group $G$ 
and connects it to the non-Abelian T-dual of the Principal Chiral Model (PCM) on $G$. 
In the past years, several generalizations have been constructed and studied. Most prominently,  $\lambda$-deformations of symmetric spaces  \cite{Sfetsos:2013wia,Hollowood:2014rla}  interpolate between  gauged WZW models with vectorial residual gauge and non-Abelian T-duals of  symmetric space $\sigma$-models. In \cite{Hollowood:2014qma} the $\lambda$-deformations were generalized to Green-Schwarz (GS) $\sigma$-models and applied to the AdS$_5\times$S$^5$ superstring. In both cases, classical integrability was easily shown using, respectively, the $\mathbb{Z}_2$ and $\mathbb{Z}_4$ grading of the spaces. Furthermore, both the non-marginal deformed (gauged) WZW models as well as the deformed GS models are known to describe consistent  type-II supergravity solutions, respectively by dressing the bosonic background with appropriate RR fields \cite{Sfetsos:2014cea,Demulder:2015lva,Itsios:2019izt} or by extracting them directly from the GS formulation \cite{Appadu:2015nfa,Borsato:2016zcf,Chervonyi:2016ajp,Borsato:2016ose2} (see also \cite{Benitez:2019oaw}  for the pure-spinor formulation of $\lambda$-AdS$_5\times$S$^5$).\footnote{In that respect, if the deformations can be embedded in supergravity by turning on RR-fields as well as fermions, they are at least to leading order, marginal.} In addition, there are constructions, starting with the work of \cite{Georgiou:2017jfi}, 
describing deformations smoothly interpolating  between exact CFTs in the UV and the IR, in full
agreement with Zamolodchikov's C-theorem \cite{Zamolo:1986}, as further was illustrated in \cite{Georgiou:2018vbb}.

In the present research we consider asymmetric $\lambda$-deformations in which the possibility of other anomaly free residual gauges, based on Lie-algebra outer automorphisms,
for the gauged WZW model was exploited \cite{Driezen:2019ykp}.\footnote{Other anomaly free gaugings utilized in the constructions of $\l$-deformations by considering tensor product CFTs, and their studies, can be found in \cite{Georgiou:2016urf,Georgiou:2017oly,Georgiou:2020eoo}.
}
In particular, when the Lie group admits such an automorphism, one can deform gauged WZW models with an asymmetric residual gauge (rather than a vectorial residual gauge) which typically describe topologically different target spacetimes \cite{Witten:1991mm,Bars:1991pt,Ginsparg:1992af}. A prominent application is a deformation of the two-dimensional Witten black hole \cite{Witten:1991yr} which was,  together with its connection to Sine-Liouville theory,  
described in \cite{Driezen:2019ykp}. 
Here we will construct  a non-trivial example utilizing an outer
automorphism, by focusing on another interesting family of CFTs, corresponding to  
the three-dimensional Horne-Horowitz black string \cite{Horne:1991gn}. This singular background is characterized by a mass, charge and dilaton profile. The gauged WZW model description in Euclidean signature is based on $SU(2)\times U(1)/U(1)_q$, 
where the gauge group $U(1)_q$  
acts axially. The parameter $q$ characterizes a linear combination of the Cartan elements and particularly  relates to the physical parameters of the black string solution. Using the asymmetric $\lambda$-formulation we can deform this theory but since the space does not admit a $\mathbb{Z}_2$-grading and $G$ is non-semisimple, classical integrability and one-loop renormalisability of the theory is not immediately ensured. However, we will show that there is an unanticipated important connection to the anisotropic XXZ (biaxial) $SU(2)$ $\lambda$-model, a particular case of the model constructed in \cite{Sfetsos:2014lla}, which is known to be integrable and renormalisable. Furthermore, an interesting feature is that in the resulting spacetimes  a $U(1)$ isometry survives after the deformation.

This paper is organized as follows: In section \ref{s:LambdaAS} we review and generalize asymmetric $\lambda$-deformations in order to include the possibility of anisotropic effects. In section \ref{s:ExamplesAxialVector} we construct the target spaces of  $\lambda$-deformations of $SU(2)\times U(1)/U(1)_q$ with a vectorial and axial residual gauge and show, in the latter case, the connection to the anisotropic $SU(2)$ $\l$-deformation. Using the surviving $U(1)$ isometry we show that the axial and vector deformations are canonically related via an Abelian T-duality. Although this is well known at the CFT level \cite{Kiritsis:1991zt,Rocek:1991ps,Giveon:1993ph} an axial-vector T-duality along the non-marginal deformation is an unseen feature.  We end this section by showing classical integrability and one-loop renormalizability.
We conclude in section \ref{conclusions} and discuss possible future directions.

\section{Asymmetric $\lambda$-deformations}
\label{s:LambdaAS}

In this section we review the work of \cite{Driezen:2019ykp}
who formulated   $\lambda$-deformations of  $G/H$ gauged WZW models with asymmetric residual gauge by generalizing the  
deformations of \cite{Sfetsos:2013wia,Hollowood:2014rla}. In addition, while the construction of \cite{Driezen:2019ykp}  was done for the isotropic case,  here we  include anisotropic effects in which the deformation is encoded in a matrix $\Lambda$ instead of a single parameter $\lambda$. 
Furthermore, we remark that these constructions  can be cast in the formalism of the usual $\l$-deformations, after a redefinition of the deformation
matrix, although integrability is more apparent in the asymmetrically gauged formulation.

Consider a general Lie group $G$  with Lie algebra generators  $t^a$,  $a \in \{ 1,2, \dots , \dim G \}$, which are normalized
 as $\Tr(t_a t_b)=\eta_{ab}$, and satisfy the  commutation relations $[t_a , t_b ] = i f_{ab}{}^c t_c$. Algebra indices will be raised and lowered using $\eta_{ab}$.
We will assume that the Lie algebra of $G$ has an automorphism $W$ acting on the generators as 
$W(t_a) = W^b{}_{a} t_b$. 
In addition, the following relations are satisfied
\be
\label{eq:AutoProp}
\Tr\big(W(t_a) W(t_b)\big) = \Tr(t_a t_b)\  ,\qq W\big([t_a,t_b]\big)= [W(t_a),W(t_b)]
= i f_{ab}{}^c W(t_c)\ . 
\ee
Consider next the asymmetrically gauged WZW action \cite{Witten:1991mm,Bars:1991pt}
\be
\begin{split}
\label{gauwzw}
S_k(g,A_\pm)= S_k(g) & + {k\ov \pi} \int d^2\s\, \Tr(A_-\del_+ g g^{-1} - B_+ g^{-1}\del_- g + A_-g B_+ g^{-1}
\\
& -\ha A_+A_- - \ha  B_+B_-)\ ,
\end{split}
\ee
where $S_k(g)$ is  the standard WZW action for the group $g$ at level $k$ given by \cite{Witten:1983ar}
\be
S_k(g) = - {k\ov 2\pi} \int d^2\s\,  \Tr(g^{-1}\del_+ g g^{-1}\del_-g) + {k\ov 12\pi} \int (g^{-1} dg)^3\ .
\ee
When $G$ is compact, the level $k$ is a positive integer, while for non-compact $G$ the  level $k$ is a positive number.
In addition, $A_\pm$ and $B_\pm$ are gauge fields which are not independent since they are built up by the same
components $A^a_\pm$. Indeed, in terms of  representation matrices we have that  $A_\pm = A_\pm^a t_a$ and  $B_\pm = B_\pm^a t_a = W(A_\pm)=A^a_\pm W(t_a)$, with $B_\pm^a = W^a{}_bA_\pm^b$. The above action is invariant under gauge transformations which in their infinitesimal
form read
\be
\d g =-\e_L g + g \e_R \ ,\quad \d A_\pm = -\del_\pm \e_L + [A_\pm,\e_L]\ , \quad \d B_\pm = -\del_\pm \e_R + [B_\pm,\e_R]\  ,
\ee
where  $\e_L  = \e^a t_a$ and $\e_R = \e^a W(t_a)$ . Notice that due to the properties of the automorphism in \eqn{eq:AutoProp}
we may replace in \eqn{gauwzw} the term $\Tr( B_+B_-)$ by $\Tr( A_+A_-)$.
In the spirit of \cite{Sfetsos:2013wia} and \cite{Driezen:2019ykp} we add to \eqn{gauwzw} the gauge invariant term
\be
\label{pcmaction}
S_{E}(\tilde g,A_\pm) = -{1\ov \pi} \int d^2\s\, E_{ab}
\Tr( t^a \tilde g^{-1} D_+ \tilde g) \Tr( t^a \tilde g^{-1} D_- \tilde g) \ ,
\ee
where $E_{ab}$ is a matrix of couplings and $D_\pm \tilde g = \del\tilde g - A_\pm \tilde g$ are the covariant derivatives for
minimal coupling. This term is on its own gauge invariant provided that $\d \tilde g = -\e_L \tilde g $. Since the gauge symmetry
acts with no fixed point in $\tilde g$ we may fix the gauge symmetry as $\tilde g = \mathbb{1}$.\footnote{We could
equally well replace \eqn{pcmaction} by an action invariant under the right transformations
 $\d \tilde g=  \tilde g \e_R$. This
would not affect our results, but it would  redefine the matrix $\Lambda$ below.}
Then after defining as in  \cite{Sfetsos:2013wia} the matrix
\be
\Lambda = k  (k\mathbb{1} + E)^{-1}\ ,
\ee
one obtains the action
\be
\label{eq:LambdaAction}
\begin{split}
S_{k,\Lambda}(g,A_\pm,W)= S_k(g) &+ {k\ov \pi} \int d^2\s\, \Tr(A_-\del_+ g g^{-1} - B_+ g^{-1}\del_- g + A_-g B_+ g^{-1}
\\
& - A_+\Lambda^{-1} A_- )\ .
\end{split}
\ee
This action is invariant under the parity-like transformation
\be
\s_+\leftrightarrow \s_-\ ,\quad g\to g^{-1}\ ,\quad A_\pm \leftrightarrow B_\mp\ ,\quad
W\to W^T\ ,\quad \Lambda \to W\Lambda^T W^T\ .
 \ee
For $W=\mathbb{1}$ the gauge fields $A_\pm$ and $B_\pm$ coincide and one would obtain, after integrating out the gauge fields, the standard $\s$-model
action  of \cite{Sfetsos:2013wia}, corresponding to $\l$-deformations of WZW models by current bilinears. Moreover, if we split the matrix $\Lambda=( \l_{G/H}, \l_H)$ and set $\l_H=\mathbb{1}_H$ and $\l_{G/H} = \l$ then we will
obtain, after integrating out the gauge fields, the $\l$-deformations of vectorially gauged coset $G/H$ CFTs \cite{Sfetsos:2013wia}.
In this latter case, but with $W$  non-trivial, one would obtain the $\l$-deformations of asymmetrically gauged coset $G/H$ CFTs of \cite{Driezen:2019ykp}.

On-shell the gauge fields satisfy the constraints
\begin{equation}
\label{eq:GaugeConstraints}
A_+ = \big( \Lambda^{-T} - D W\big)^{-1} \partial_+ g g^{-1} , \qquad
A_- =  \big(D^{-1}  - W \Lambda^{-1}\big)^{-1} g^{-1} \partial_- g\  ,
\end{equation}
where  $D$ represents the adjoint action as $D (t_a) = D^b{}_a t_b= g t_a g^{-1}$ and thus $D_{ab}=\Tr(t_a gt_b g^{-1})$. 
The resulting effective action valid at large $k$, but
exactly in the matrix $\Lambda$, is obtained by substituting  \eqn{eq:GaugeConstraints} into  \eqn{eq:LambdaAction}
\begin{equation}
 \label{eq:LambdaActionEffective}
S_{k,\Lambda} (g, W) =  S_{k}(g)  + \frac{k}{\pi} \int d^2 \sigma\,
J^a_+ \big( W\Lambda^{-1}-D^{-1} \big)^{-1}_{ab} J_-^b \ ,
\end{equation}
where $J_+^a= - i\, \Tr(t^a \del_+ g g^{-1})$ and $J_-^a= - i\, \Tr(t^a g^{-1} \del_+ g )$.
This action is accompanied with a non-constant dilaton profile coming from the elimination of gauge fields when performed in the path integral
\begin{equation}
\label{diil}
e^{-2\Phi} = e^{-2\Phi_0 }  \det \eta \det \big(\Lambda^{-1} -  W^TD^T \big) \ ,
\end{equation}
with $\Phi_0$ constant. In general this contribution is   important if one attempts to embed these models to supergravity. In this work, it plays a r\^ole in discovering the part of the
diffeomorphism needed to compute the $\beta$-functions as we will see explicitly below.
Note that, as in the anisotropic PCM,  the action \eqn{eq:LambdaActionEffective} is expected to be integrable only for special choices of the matrix $\L_{ab}$.

The above action, as in the original $\lambda$-deformations, has the symmetry  \cite{Itsios:2014lca,Sfetsos:2014jfa}
\begin{equation}
\label{eq:LambdaDisSymm}
g \rightarrow g^{-1}, \qquad \Lambda \rightarrow \Lambda^{-1} , \qquad k \rightarrow -k\  .
\end{equation}
In the case with $\Lambda=( \l_{G/H}, \mathbb{1}_H)$ there is also a residual  $\dim H$ asymmetrical gauge invariance acting as
\begin{equation} \label{eq:ResidualGauge}
\begin{aligned}
g \rightarrow h^{-1} g \widetilde{h}\ , ~~~~A_\pm^{(H)} \rightarrow h^{-1} (A_\pm^{(H)} - \partial_\pm ) h\ , ~~~~
 A_\pm^{(G/H)} \rightarrow h^{-1} A_\pm^{(G/H)} h\  ,
\end{aligned}
\end{equation}
with $h = e^X$ and $\widetilde{h} = e^{W(X)}$ (connected to the identity) for $X$ in the Lie subalgebra of $H \subset G$.
Hence, the fields $A_\pm^{(H)}$ are still genuine (but non-propagating) gauge fields, while $A_\pm^{(G/H)}$ are auxiliary.
That means that we should gauge fix $\dim H$ parameters in the group element $g$ in the effective action \eqn{eq:LambdaActionEffective}.

 An important remark is that the effect of the automorphism $W$ is only non-trivial when $W$ is an outer automorphism of $G$.
For instance, let us assume that it is possible to write $W(t_a) = w t_a w^{-1}$ for some $w$. Then, for any $\Lambda$ eq.~\eqref{eq:LambdaAction} can be rewritten as
$
S_{k,\Lambda} (g , A_\pm, W) = S_{k,\Lambda} (gw, A_\pm , \mathbb{1})
$,
similar as in the undeformed case. When the automorphism $W$ is inner, that is when $w\in G$, all that one obtains is a trivial field redefinition from $g\in G$ to $gw \in G$ so that the construction is the same as in the original $W=\mathbb{1}$ case, that is the vector gauging.
When the automorphism $W$ is outer, such a field redefinition is not possible, and the asymmetrical gauging will deform target spaces
which are topologically different from the usual vectorially gauged cases \cite{Witten:1991mm,Bars:1991pt,Ginsparg:1992af}. In particular, when  the matrix entries
for $W$ are different than unity in the directions of a subgroup $H \subset G$ one can  systematically construct   deformations of the asymmetric cosets $G/H$ CFTs.

Classical integrability as well as one-loop renormalisability of the asymmetrical isotropic $\lambda$-deformations was demonstrated in \cite{Driezen:2019ykp} in the cases that $\Lambda = \lambda \mathbb{1}_G$ and $\Lambda=( \l \mathbb{1}_{G/H} , \mathbb{1}_H)$ assuming for the latter that the Lie subalgebra $Lie(H)$ underlies a $\mathbb{Z}_2$ grading for $Lie(G)$.\footnote{In \cite{Driezen:2019ykp} it was also illustrated how non-trivial outer automorphisms can be incorporated into $\lambda$-deformations on semi-symmetric spaces, in which case $Lie(G)$ admits a $Z_4$ grading, without destroying classical integrability, by generalizing the construction of \cite{Hollowood:2014qma}.}

The form of the action \eqn{eq:LambdaActionEffective} and the dilaton \eqn{diil}
suggest that the effect of an automorphism  can be absorbed in the deformation matrix $\Lambda$, by redefining $\Lambda\to \Lambda W$ and using  that $W$ satisfies $W \eta W^T = \eta$. This fact enables one to map these models to the 
$\l$-deformed models of \cite{Sfetsos:2013wia,Georgiou:2016urf} with a general coupling matrix and therefore adds a class of integrable deformed models constructed from non-trivial outer automorphisms.

\section{$\lambda$-deformations of  $SU(2)\times U(1)/U(1)_q$} \label{s:ExamplesAxialVector}

In this section we explicitly construct the two possible $\l$-deformations of the $SU(2)\times U(1)/U(1)_q$ coset CFT which have an axial and vectorial residual gauge.
Although these deformations are not marginal, we will show that the axial-vector T-duality persists along the deformation line, and hence that they are canonically equivalent according to \cite{Curtright:1994be,Alvarez:1994wj}.
One of  the deformations--namely the one corresponding to the axial gauging--corresponds to 
the deformation of the Euclidean black string solution of \cite{Horne:1991gn}. Although $SU(2)\times U(1)$ is non-semisimple and the subalgebra $Lie(U(1)_q)$ does not realize a $\mathbb{Z}_2$ grading of $Lie(SU(2)\times U(1))$, we still demonstrate classical integrability of the deformed theories and we compute the one-loop $RG$-flow equations of parameters. Finally, we show that remarkably the axial deformation can be described as the particular biaxial anisotropic $SU(2)$ $\lambda$-deformation constructed in \cite{Sfetsos:2014lla}.

We will parametrize a group element $g \in G$ by Euler angles as
\begin{equation}
\label{eq:GroupEl}
g  = g_{\text{SU(2)}} \cdot g_{\text{U(1)}}  \ ,
\end{equation}
where
\begin{equation}
\begin{split}
& g_{\text{SU(2)}} =  e^{\frac{i (\phi_1 - \phi_2)  }{2} \sigma_3}  e^{i \omega \sigma_1}   e^{\frac{i (\phi_1 + \phi_2) }{2}\sigma_3} =  \begin{pmatrix}
e^{i \phi_1} \cos\omega & i e^{- i \phi_2} \sin\omega \\ i e^{i \phi_2} \sin\omega & e^{-i \phi_1}\cos\omega
\end{pmatrix}\ ,
\\
& g_{\text{U(1)}} = e^{i x  \sigma_0}\ ,
\end{split}
\end{equation}
with $\sigma_i$ being the usual Pauli matrices and $\sigma_0 = \mathbb{1}_2$. We will utilize the corresponding Lie algebra in the following basis
\begin{equation}
\label{eq:generatorsBlackString}
\begin{aligned}
t = \frac{1}{\sqrt{2}} \bigg\{ \sigma_1, \sigma_2 , \frac{- q \sigma_3 + \sigma_0}{\sqrt{1+q^2}} , \frac{\sigma_3 + q \sigma_0}{\sqrt{1+ q^2}} \bigg\}\ ,
\end{aligned}
\end{equation}
where the parameter $q \in \mathbb{R}$ defines the relative weights in   mixing   $\sigma_0$ and $\sigma_3$
in the Cartan subalgebra, and the overall factor is such that $\Tr (t_a t_b) = \delta_{ab}$.
In this basis the non-vanishing and $q$-dependent Lie algebra structure constants are
\be
\label{struu}
f_{123}= -{\sqrt{2}\ q\ov \sqrt{1+q^2}}\ ,\qq f_{124}= {\sqrt{2} \ov \sqrt{1+q^2}}\ ,
\ee
the rest of the components follow from antisymmetry.
We will construct $\lambda$-deformations with  both the vector and the axial-type by gauging
the subgroup $H$ generated by $t_4$. Hence 
we take the deformation matrix in the basis \eqn{eq:generatorsBlackString} to be diagonal as
\be
\label{matrrlab}
(\Lambda_{ab}) =\diag(\l,\l,\l_3 ,\l_4) = \diag(\l,\l,\l,1) \ .
\ee
The axial-type then deforms directly the Euclidean black string \cite{Horne:1991gn} for which we recall that the physical properties depend on $q$. At this point, the parameter $q$  simply determines the relative weights of  gauging a subgroup action generated by $\sigma_0$ vs.~$\sigma_3$. We denote this subgroup as $U(1)_q$. 

 For the (original) vector $\lambda$-model we must take $W = \mathbf{1}_4$. 
 The infinitesimal residual vector gauge  transformation $\delta_V g =i \epsilon \left[t_4 , g\right] $ acts explicitly as
\begin{equation}
\delta_V x = 0\ , \quad \delta_V \phi_1 =0\ ,\quad \delta_V\phi_2 = - \frac{2\epsilon}{\sqrt{2(1+q^2)}} \ , \quad \delta_V \omega = 0 ,
\end{equation}
and thus taking $\phi_2 = 0$ in \eqref{eq:GroupEl} fixes the gauge completely.

For the axial $\lambda$-model the automorphism $W$ should act on the subgroup $U(1)_q$ as $W(t_4) = -t_4$. To ensure the properties \eqn{eq:AutoProp} of $W$ on the full $SU(2)\times U(1)$ algebra, it should act as
\begin{equation}
\label{eq:Automorphism}
W \{ t_1, t_2 , t_3, t_4 \} = \{  t_1 , - t_2 , - t_3, - t_4\}\ .
\end{equation}
We note  that this action can not be written as $W(t_a) = w t_a w^{-1}$ for some  two-dimensional matrix $w$ and thus $W$ is an outer automorphism. Other choices, such as flipping the sign of $t_1$ instead of $t_2$ are connected to \eqref{eq:Automorphism} by an inner automorphism $W'(t)_a = w W(t)_a w^{-1}$ with $w = \text{diag}(1,-1) \in SU(2)\times U(1)$. Indeed, this $w$ can be obtained from \eqn{eq:GroupEl} by setting $\omega\! =\! 0$, $\phi_1\!  =\! \pi/2$,  $x\! =\! -\pi/2$. Hence this apparent ambiguity amounts to a field redefinition between the resulting $\sigma$-models.
The infinitesimal residual gauge transformation  $\delta_A g = i \epsilon (g t_4 + t_4 g )$ transforms both
$x$ and $\phi_1$ in \eqref{eq:GroupEl}
\begin{equation}
\label{gaaa}
\delta_A x = \frac{2 q \epsilon}{\sqrt{2(1+q^2)}}\ , \quad \delta_A \phi_1 = \frac{2\epsilon}{\sqrt{2(1+q^2)}}\ , \quad \delta_A \phi_2 = 0 \ , \quad \delta_A \omega = 0\ .
\end{equation}
Choosing $x = 0$ completely fixes the axial gauge freedom.

There are two particular limits concerning the parameter $q$ for which these models are known. When $q\rightarrow 0$, which we will refer to as the coset limit, $t_4 \rightarrow \sigma_3$. In that case both the axial and vector gaugings would produce, up to a trivial field redefinition, the same $SU(2)/U(1) \times U(1)$ (deformed) background since $SU(2)$ has no outer automorphisms. When $q\rightarrow \infty$, which we will refer to as the group limit, $t_4 \rightarrow \sigma_0$ and the axial gauging would lead to the $\lambda$-deformed $SU(2)$ group manifold.

Let us finally point out that one may cast the above deformation in terms of the untwisted basis
 \begin{equation}
 \label{eq:generatorsun}
\tilde t = \frac{1}{\sqrt{2}} \big\{ \sigma_1, \sigma_2 , \sigma_3,\sigma_0 \big\}\ ,
\end{equation}
but with the deformation matrix $\tilde{\Lambda}$ in \eqn{eq:LambdaActionEffective} being non-diagonal.
For both the vector and the axial cases  this is easily found to be
\be
\begin{split}
\label{matrrlabNONv}
(\tilde{\Lambda}_{ab}) & =   \left(
  \begin{array}{cccc}
   \l & 0 & 0 & 0 \\
    0 & \l & 0 & 0 \\
    0 & 0 &  {q^2 \l_3 + \l_4 \ov  1+q^2} &  {q(\l_4-\l_3)\ov  1+q^2} \\
    0 & 0 &  {q(\l_4-\l_3)\ov  1+q^2} & {\l_3 + q^2\l_4 \ov  1+q^2}\\
  \end{array}
\right)  =  \left(
  \begin{array}{cccc}
   \l & 0 & 0 & 0 \\
    0 & \l & 0 & 0 \\
    0 & 0 & { 1+ q^2 \l\ov  1+q^2}  &  {q(1-\l)\ov  1+q^2} \\
    0 & 0 &  {q(1-\l)\ov  1+q^2} & {\l + q^2\ov  1+q^2} \\
  \end{array}
\right)\ ,
\end{split}
\ee
where we took $\lambda_3 = \lambda$ and $\lambda_4 = 1$ as in \eqref{matrrlab}.

\subsection{Deformed $\sigma$-model geometry}

\subsubsection{Vectorial gauge symmetry}

Using the above, the vectorial $\lambda$-deformation of the $SU(2)_k\times U(1)/U(1)_q$ coset CFT as
 derived from \eqref{eq:LambdaActionEffective} is found to have the metric
\begin{equation}
\label{eq:vectorLambda}
\begin{split}
 ds^2 = k\bigg(& {1+\l\ov 1-\l}(1+q^2)dx^2 + {1-\lambda\ov 1+\l}\Big(d\om^2+\cot^2\om  (d\phi_1+q dx)^2\Big)
 \\
 & + {4\l\ov 1-\l^2}\Big(\cos\phi_1 d\om + \sin\phi_1 \cot\om (d\phi_1+q dx)\Big)^2\bigg)\ ,
\end{split}
\end{equation}
and dilaton
\be
\label{eq:vectorLambda1}
e^{-2\Phi} = e^{-2\Phi_0} \sin^2\omega\ ,
\ee
whereas the antisymmetric tensor is vanishing.
The undeformed $\lambda= 0$ point gives the metric
\begin{equation}
d s^2 = k \left((1+q^2) dx^2 +  d\omega^2   + \cot^2\omega ( d\phi_1 + q dx)^2  \,  \right) .
\end{equation}
In this case the $q$-dependence can be undone by an obvious shift of the $\phi_1$ coordinate\footnote{Here we do not pay particular attention to global issues related to the range of values of the various variables.} and  a rescaling of the free boson $x$. 
The non-trivial part of the metric was found in \cite{Bardacki:1990wj} and corresponds to the usual $SU(2)/U(1)$ parafermionic CFT \cite{Fateev:1985mm}. Its interpretation of its Minkowski  analytic continuation as a  two-dimensional black hole and the importance of the dilaton factor \eqn{eq:vectorLambda1} was discussed in \cite{Witten:1991yr}.
For generic $\lambda$, however, the $q$-dependence can not be undone as  the isometry in $\phi_1$ is destroyed by the deformation. The isometry in the $x$-direction is on the other hand preserved which, as we will see below, allows us to perform a T-duality transformation.\footnote{We have verified that the Killing vector equations admit only the isometry along $x$.}  Furthermore, one can readily verify that the $q\rightarrow 0 $ limit gives the $\lambda$-deformed $SU(2)/U(1)$ background of  \cite{Sfetsos:2013wia} times an additional $U(1)$ factor in $x$. As a side note, one  can also verify that the non-perturbative symmetry \eqref{eq:LambdaDisSymm} (where $g\rightarrow g^{-1}$ sends $\omega \rightarrow - \omega$, $\phi_1 \rightarrow - \phi_1$, $x\rightarrow - x$ and $\phi_2 \rightarrow \phi_2$) indeed holds in the background geometry. In addition, there
is the discrete symmetry
\be
\label{qx}
q\to -q\ ,\qq x\to -x\ .
\ee
Physical quantities, such as the $\beta$-function
equations and operator anomalous dimensions, should respect this discrete  symmetry as well
as \eqn{eq:LambdaDisSymm}.

Finally, note the zoom-in limit near the singularity at $\om=0$. This strong coupling limit makes
sense if we also take $\l$ near one. Specifically, let, as in  \cite{Sfetsos:2013wia},
\be
\label{nonabellimit}
\phi_1= {x_1\ov 2k}\ ,\qquad \om = {x_2\ov 2k}\ , \qquad x={y\ov 2k}\ ,\qquad \l= 1-{\kappa^2\ov k}\ ,
\ee
and take the limit $k\to \infty$. We find that
\be
\label{dsnonabv}
ds^2 = {1+q^2\ov 2\k^2} dy^2 +
{ \k^2\ov 2} {(dx_1+q dy)^2\ov   x_2^2} + {1\ov 2\k^2}\left(dx_2
+ x_1 {dx_1+q dy\ov x_2}\right)^2 \ ,
\ee
and that
\be
\label{dsnonabvdil}
e^{-2\Phi} = e^{-2\Phi_0} x_2^2\ ,
\ee
where we have appropriately shifted $\Phi_0$ in order to absorb a $k$-dependent constant and
make sense of the limiting procedure. According to the general results of \cite{Sfetsos:2013wia} the above background corresponds to the non-Abelian T-dual   of the coset $SU(2)\times U(1)/ U(1)_q$ (in which the right acting $U(1)_q$ is gauged) performed with respect to the left acting $(SU(2)\times U(1))_L$ isometry. In addition, we have have checked that this is a genuine three-dimensional geometry and not of the direct product type.

\subsubsection{Axial gauge symmetry and anisotropic $SU(2)$}

We now turn to the case of the axial gauging.  Using the above, the background of the axial $\lambda$-deformation of $SU(2)_k\times U(1)/U(1)_q$ as derived from \eqref{eq:LambdaActionEffective} has  the metric
\ba
\label{eq:axialLambda}
&& ds^2 = k\, \frac{ 1+\lambda}{1-\lambda}\, {1\ov \D}
\nonumber
\\
&&\quad  \Big( q^2 (1-\lambda)^2 \cos^2\omega \ d\phi_1^2 + (1+q^2) \left(  (1-\lambda)^2 \cos^2\phi_2 +(1+\lambda)^2 \sin^2\phi_2   \right) \sin^2\omega \ d\phi_2^2
\nonumber
\\
&& \quad +  \left ( q^2 (1-\lambda)^2 + ( (1-\lambda)^2 + 4 \lambda (1+q^2) \cos^2\phi_2   ) \cos^2\omega \right) d\omega^2
\\
&&\quad -2\lambda (1+q^2) \sin 2\phi_2 \sin 2\omega \ d\phi_2 d\omega \Big)\, ,
\nonumber
\ea
where
\be
\label{eq:axialLambda1}
\Delta = (1+\lambda)^2 \cos^2\omega + q^2 \big( (1+\lambda)^2
- 4 \lambda \sin^2\omega \cos^2\phi_2 \big)\ .
\ee
Unlike the vectorial case, there is a non-vanishing antisymmetric tensor given by
\be
\label{eq:axialLambda2}
\begin{split}
B =& - \frac{k}{2} \cos 2\omega \ d\phi_1 \wedge d\phi_2
\\
&
- {k\ov \D}  \Big( \big((1+\lambda)^2 + 4\lambda\, q^2
 \cos^2\phi_2  \big) \cos^2\omega \sin^2 \omega \ d\phi_1 \wedge d\phi_2
\\
&\qquad + \l\,  q^2  \sin 2\omega \sin 2\phi_2 \ d\phi_1 \wedge d\omega  \Big) \ ,
\end{split}
\ee
where the first line arises from a specific gauge choice for the two-form field coming from the WZ term on $SU(2)$.
Finally, the dilaton is
\be
\label{eq:axialLambda3}
e^{-2\Phi} =  e^{-2\Phi_0} \Delta\ .
\ee
Possible singularities of the above background may arise from locations where $\D$ vanishes. It turns out that, there are no such points and  therefore our background is not singular.

\no
In the undeformed  limit the above reduces to the CFT background
\be
\label{limitaxial}
\begin{split}
& ds^2 =k \Big(d\om^2 + {1\ov q^2+\cos^2\om}(q^2 \cos^2\om\, d\phi_1^2+ (1+q^2)\sin^2\om\, d\phi_2^2 )\Big)\  ,
\\
& B = k {q^2\sin^2\om\ov  q^2+\cos^2\om}\, d\phi_1 \wedge d\phi_2 \ ,
\\
& e^{-2\Phi} =  e^{-2\Phi_0}  ( q^2 + \cos^2\omega )\  ,
\end{split}
\ee
where we have neglected in $B$ the term $\displaystyle -{k\ov 2} d\phi_1 \wedge d\phi_2$. This is the Euclidean analytic
continuation of the black string solution of \cite{Horne:1991gn}. Hence, the solution \eqn{eq:axialLambda}-\eqn{eq:axialLambda2} is the $\l$-deformation of the above mentioned CFT. 

An important feature of the axial deformation is that it can be recast in terms of the
three-parameter anisotropic $SU(2)$ $\l$-deformation of \cite{Sfetsos:2014lla} in the particular case when two of the parameters are equal, i.e.~the biaxial or XXZ case.
This is seen by computing \eqn{eq:LambdaActionEffective} and \eqn{diil} for $SU(2)$ in the usual 
basis $t={1\ov \sqrt{2}}(\s_2,\s_2,\s_3)$ and $W=\mathbb{1}_3$ with deformation matrix 
\begin{equation} 
\label{eq:LambdaAni}
(\Lambda_{ab}) =  \text{diag} \left(\lambda , \lambda , \frac{1+\lambda + 2 {q}^2 \lambda}{1+\lambda + 2 {q}^2}
\right)\, .
\end{equation}
We find that the metric, antisymmetric tensor and dilaton  coincides with the axial background upon the field redefinitions $\phi_1 \leftrightarrow \phi_2$ and $\omega \rightarrow \omega - \frac{\pi}{2}$. Due to the form of the above matrix, this is an anisotropic $SU(2)$ deformation of XXZ type, a particular 
case of the fully anisotropic model  $\Lambda = \text{diag}(\lambda_1, \lambda_2 , \lambda_3)$ constructed in  \cite{Sfetsos:2014lla}. 

Finally, let
\begin{equation}
\label{nonabellimit2}
\phi_1 =\frac{2 \tilde{y}}{q}\ , \qquad \phi_2 =\frac{x_1}{2k}\ ,
\qquad  \omega =\frac{x_2}{2k} - \frac{\pi}{2}\ , \qquad \lambda = 1 - \frac{\kappa^2}{k}\ ,
\end{equation}
which is analogous to  \eqn{nonabellimit}. Then in the limit $k\rightarrow \infty$,
the background \eqn{eq:axialLambda1}-\eqn{eq:axialLambda3}, becomes
\begin{equation}
\begin{split}
& ds^2 =\Gamma^{-1} \bigg(2 \kappa^2 x_2^2 d\tilde{y}^2 + {1+q^2\ov 2\k^2} (x_1 dx_1 + x_2 dx_2)^2  +{(1+q^2) \kappa^2\ov 2} dx_1^2  +  {q^2 \kappa^2\ov 2} dx_2^2\bigg)\ ,
\\
& B = \G^{-1}
\bigg({1\ov 2 q} \big(x_2^2 - q^2(x_1^2 - x_2^2 + \kappa^4)\big)\, dx_1  -  q x_1 x_2\, dx_2 \biggl)\wedge d\tilde y \ ,
  \\
& e^{-2\Phi} = e^{-2\Phi_0}\G\ ,
\end{split}
\end{equation}
with the function
\be
\G =q^2 x_1^2 + (1+q^2)  x_2^2 + q^2 \kappa^4 ,
\ee
and where we have performed a constant shift in the dilaton to take a sensible limit in $k\rightarrow \infty$.
Again this background is non-singular. Whilst in general \cite{Driezen:2019ykp} the limit $k\rightarrow \infty$ of  asymmetrical $\lambda$ models does not have a non-Abelian T-dual interpretation,  we will see in what follows that the deformed Euclidean black string in this limit is the Abelian T-dual to \eqn{dsnonabv} and \eqn{dsnonabvdil}, a feature that goes
beyond the non-Abelian limits.

\subsection{Relating the vector and axially deformed theories via T-duality}

We would like to point out the striking connection between the deformed axial and vector theories via Abelian T-duality.
This is well known in the  undeformed coset CFTs  at $\lambda = 0$ \cite{Kiritsis:1991zt,Rocek:1991ps,Giveon:1993ph}, but nevertheless it persists along the full (non-marginal) deformation line.  We first make the field renamings
$x \rightarrow \theta/k$, $\phi_1 \rightarrow \phi_2$ and $\omega \rightarrow \omega - \pi/2$ 
in the vector background  \eqref{eq:vectorLambda}  and \eqn{eq:vectorLambda1} as well as
$\phi_1 \rightarrow \widetilde{\theta}/q$ 
in the axial background \eqref{eq:axialLambda}-\eqn{eq:axialLambda3}.
Starting from the redefined vector background one can perform a T-duality along the  isometric coordinate $\theta$ using the well-known Buscher procedure.
Then, we produce precisely the (redefined) axial background \eqref{eq:axialLambda}-\eqn{eq:axialLambda3}. Hence, for generic $\lambda$  we have an Abelian axial-vector T-duality between  field theories which are non-conformal at the bosonic level. This means that the two deformed backgrounds are  canonically equivalent,
implying for instance that one may verify equally well classical integrability for either one of them. A similar comment applies for the $\beta$-function equations, i.e.~we
may use either background to compute the $\beta$-functions for the couplings $\l$ and $q$. Below in
fig.~\ref{f:ABTD} we have schematically encoded
the interplay of T-duality and the $\l$-deformations in our example.
\begin{figure}[H]
\centering
\includegraphics[scale=.26]{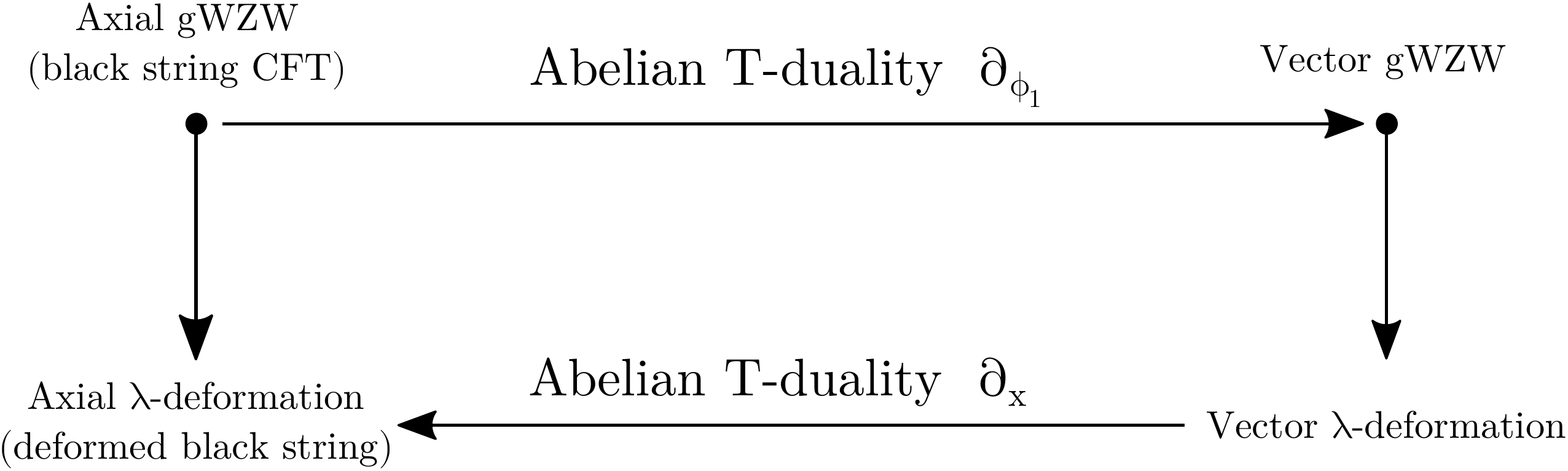}
\caption{Web of T-dualities and $\l$-deformations.}
\label{f:ABTD}
\end{figure}

\subsection{Classical integrability and the Lax pair}

The $(SU(2)\times U(1)) /U(1)_q$ coset does not have a $Z_2$ grading. This is apparent from the fact that the structure constants with solely coset generators are non-vanishing,
i.e.~$f_{123}$ in \eqn{struu}.
Hence, the general proof of integrability by encoding the classical equations of motion in terms of a Lax pair given for any $\l$-deformed symmetric space in \cite{Hollowood:2014rla} and in \cite{Driezen:2019ykp} (for non-trivial outer automorphisms) does not apply in our case.
Instead, we have to investigate integrability issues for the backgrounds corresponding to the vector and axial gauge symmetries on their own.
Since these are related by T-duality and the latter preserves integrability we may focus on either the
deformation based on the vector gauging or on that based on  the axial one. We found it more conveniently to work with
the axial case and in particular the formulation in which the automorphism is given by \eqn{eq:Automorphism} and the deformation matrix  by \eqn{matrrlab}. 
Furthermore, even though our model is equivalent to the biaxial $\l$-deformed anisotropic $SU(2)$ model of \cite{Sfetsos:2014lla} which is integrable, we present here the explicit derivation of the Lax pair in the axial formulation having in mind future applications.

The equations of motion for the general asymmetrical $\l$-deformed model \eqn{eq:LambdaAction} can be cast in terms of the gauge fields satisfying \eqn{eq:GaugeConstraints} as
\be
\begin{split}
\label{eomAinitial}
& \del_+ A_- - \Lambda^{-T} \del_- ( A_+) = [\Lambda^{-T} A_+,A_-]\ ,
\\
& \Lambda^{-1} \del_+(A_-)-\del_-A_+=[A_+,\Lambda^{-1}A_-]\ ,
\end{split}
\ee
where we have used the fact that $W$ is an automorphism.  Note that they have the same form as the corresponding equations of motion in the $W=\mathbb{1}$ case \cite{Sfetsos:2014lla}. 
Nevertheless, the dependence on the particular automorphism is hidden in the on-shell expressions for the gauge fields in \eqn{eq:GaugeConstraints}.
In order to specialize to the coset case we let  the matrix $\Lambda=(\l \mathbb{1}_{G/H} ,\mathbb{1}_H )$. Next we also split the gauge fields as
$A_\pm =\big(A_\pm^{(G/H)}, A_\pm^{(H)}\big)$. Then after projecting to the subgroup
and to the coset we cast these equations as
(two of the projected equations are actually equivalent)
 \be
\label{eq:eom}
 \begin{split}
 & \partial_- A^{(H)}_+ - \partial_+ A^{(H)}_- + \left[ A^{(H)}_+ , A^{(H)}_- \right] + \lambda^{-1} \left[ A^{(G/H)}_+ , A^{(G/H)}_- \right] = 0\ ,
 \\
 & \partial_- A^{(G/H)}_+ - \lambda^{-1} \partial_+ A^{(G/H)}_- + \lambda^{-1} \left[A^{(H)}_+, A^{(G/H)}_- \right]
 \\
 & \qq\qq\qq + \left[ A^{(G/H)}_+ , A^{(H)}_- \right] +  \lambda^{-1}\left[ A^{(G/H)}_+  , A^{(G/H)}_- \right]     = 0\ ,
 \\
&  \lambda^{-1} \partial_- A^{(G/H)}_+ -  \partial_+ A^{(G/H)}_- +  \left[A^{(H)}_+, A^{(G/H)}_- \right]
\\
&\qq\qq\qq  + \lambda^{-1} \left[ A^{(G/H)}_+ , A^{(H)}_- \right] +  \lambda^{-1}\left[ A^{(G/H)}_+  , A^{(G/H)}_- \right]     = 0\  ,
 \end{split}
 \ee
where the last terms of the second and third equations are not present in the case of symmetric spaces. In the case at hand it seems that there are seven equations in total.
However, it turns out that from the on-shell constraints for the gauge fields \eqn{eq:GaugeConstraints} one can derive the following relation
\be
2 q\lambda\, A_\pm^4 + (1+\lambda) A_\pm^3 = 0 \ ,
\label{eq:identity}
\ee
where we have used the basis in \eqn{eq:generatorsBlackString}.
By means of this relation one may readily verify that the first equation in \eqn{eq:eom} is already captured by the projection to $t_3$ of the sum of the second and the third equations.
Hence, there is a total of six independent equations in terms of six independent gauge fields $A_\pm^\alpha$, $\alpha=1,2,3$, which determine the classical evolution of the fields $\{\omega, \phi_1, \phi_2\}$.

Next, we try to find the Lax connection representing \eqref{eq:eom} by making the following ansatz based on symmetry arguments
\begin{equation}
\begin{aligned}
{\cal L}_\pm (z) &= a_\pm A_\pm^4  t_4 + b_\pm \big( A_\pm^1 t_1 + A_\pm^2 t_2 \big) + c_\pm A_\pm^3 t_3 ,
\end{aligned}
\end{equation}
with $a_\pm, b_\pm$ and $c_\pm $  unknown constants. In order to generate conserved charges, at least one of the unknown constants should depend on an arbitrary complex parameter \cite{Zakharov:1973pp}. They will be determined by enforcing that  the Lax flatness condition
\begin{equation}
\label{eq:LambdaBSFlatLax}
\partial_+ {\cal L}_- (z) - \partial_- {\cal L}_+(z) - \left[ {\cal L}_+(z) , {\cal L}_- (z) \right] = 0\ ,
\end{equation}
is solved by  \eqref{eq:eom} and the relation \eqn{eq:identity}.
After substituting the solutions for the derivatives of the gauge fields $A_\pm^\alpha$ found from \eqref{eq:eom} into \eqref{eq:LambdaBSFlatLax} we remarkably find only four independent equations for the six unknown coefficients, given by
\be
\label{eq:int}
\begin{split}
& \frac{q}{\sqrt{1+q^2}} \big( c_+ + c_- - (1+\lambda) b_+ b_- \big) = 0\ ,
 \\
& \frac{1}{\sqrt{1+q^2}}\left(  a_+ + a_- - 2 \lambda b_+ b_-  \right) = 0\ ,
\\
&  \frac{q}{\sqrt{1+q^2}} \big(b_+ + b_- - (1+\lambda)  c_+ b_-\big) + \frac{(1+\lambda)^2 (1- a_+ ) b_-}{2q\sqrt{1+q^2}\, \lambda}  = 0 \ ,
 \\
&  \frac{q}{\sqrt{1+q^2}} \big(b_+ + b_- - (1+\lambda)  b_+ c_-\big) + \frac{(1+\lambda)^2 (1- a_-) b_+}{2q\sqrt{1+q^2}\, \lambda}  = 0\  ,
\end{split}
\ee
 and where the overall factors in  $q$ have been kept in order to be able to take the extreme limiting
cases below. We thus  clearly have a  \textit{two}-parameter redundancy in the system.
In the limiting cases, however, this reduces to the ordinary one-parameter redundancy.
Indeed, for $q \rightarrow 0 $ the  $t_3$-direction  decouples because of \eqref{eq:identity} and
 the first equation of \eqref{eq:int} is non-existing. Hence we have three equations for the four unknowns $a_\pm, b_\pm$ which one can verify  is solved by the symmetric coset solution of \cite{Hollowood:2014rla}
\begin{equation}
\label{eq:CosetSol}
\begin{aligned}
 q \rightarrow 0 : \qquad {\cal L}_\pm (z) = A_\pm^4 t_4  + z^{\pm 1} \lambda^{-1/2}
 \big( A_\pm^1 t_1 + A_\pm^2 t_2 \big) \ ,
 \end{aligned}
 \end{equation}
 where  $z \in \mathbb{C}$ is the only arbitrary spectral parameter. 
Similarly,  for  the limit $q \rightarrow \infty$ the  $t_4$-direction decouples and the second equation of \eqref{eq:int} is non-existing, so that we have three equations for the four unknowns $b_\pm, c_\pm$ which are solved  by the group solution \cite{Hollowood:2014rla} 
 \begin{equation}
 \label{eq:GroupSol}
 \begin{aligned}
 q \rightarrow \infty : \qquad &{\cal L}_\pm (z) =  \frac{2}{1+\lambda}\frac{1}{1\mp z} A^{\alpha}_\pm t_\alpha \ , \qq \alpha = 1,2,3 \ .
 \end{aligned}
 \end{equation}
For general values of the parameter $q$, on the other hand,  we have a two parameter family of solutions
of \eqn{eq:int}. The solution is conveniently presented by first defining the functions
\be
\begin{split}
& a(\xi, \zeta, \epsilon) \equiv \frac{(1+\lambda)^2 \xi (1-\zeta ) +2 q^2 \lambda (\xi-\zeta ) (1+\xi) }{(1+\lambda)^2 (\epsilon - \zeta) \xi  }\ ,
\\
& b(\xi, \zeta, \epsilon) \equiv  \sqrt{\frac{(1+\lambda)^2 \xi + q^2 \lambda (1+\xi)^2 }{(1+q^2) \lambda (1+\lambda)^2 \xi^2  }}\ ,
\\
& c(\xi, \zeta, \epsilon) \equiv  \frac{(1+\lambda)^2 \xi (\epsilon - 1) + 2 q^2 \lambda (\epsilon - \xi) (1+\xi) }{2q^2 \lambda (1+\lambda) (\epsilon - \zeta) \xi} \ ,
\end{split}
\ee
where $\xi$, $\zeta$ and  $\epsilon$ are complex parameters which are related by
the constraint
  \begin{equation}
  \label{eq:ConstraintSols}
\begin{aligned}
& \xi (\epsilon - \zeta) +  {1+q^2\ov 2}  \xi (\epsilon +1 ) (\zeta - 1)
\\
&\quad + {\l\, q^2 \ov (1+\l)^2} \Big( (\xi + 1)^2 (\epsilon - \zeta) +   (1+q^2)
(\zeta - \xi) (\epsilon+1) ( \xi + 1)\Big)   = 0\  ,
\end{aligned}
 \end{equation}
 allowing for two independent parameters among them as advertised.\footnote{Of course the above expression \eqref{eq:ConstraintSols} can be easily solved for $\e$. However, the  solution
 is not particularly illuminating so that we will not present it explicitly.}
The various coefficients in the Lax pair turn out to be
\begin{equation}
 a_\pm = a(\xi^{\pm 1},\zeta^{\pm 1},\epsilon^{\pm 1})\ , \qquad b_\pm = b(\xi^{\pm 1},\zeta^{\pm 1},\epsilon^{\pm 1})\ ,   \qquad c_\pm = c(\xi^{\pm 1},\zeta^{\pm 1},\epsilon^{\pm 1})\  .
\end{equation}
One can verify that this solution reduces to the coset case \eqref{eq:CosetSol} in the $q\rightarrow 0 $ limit (with $c_\pm$ decoupled) after the redefinition $\xi = z^2$ and to the group solution \eqref{eq:GroupSol} in the  $q\rightarrow \infty$ limit (with $a_\pm$ decoupled) after the redefinition $\displaystyle \xi = \frac{1-z}{1+z}$.

We have represented  the equations of motion on a non-symmetric space in terms of a flat Lax connection which depends on  two arbitrary (spectral) parameters. This suggests an excessive generation of conserved  charges.  As far as we are aware, there are no such examples known in the literature; at least in the  landscape of integrable deformed $\sigma$-models.  
However,  the origin of the second redundancy becomes clear when  we recast the axial model as the anisotropic XXZ $SU(2)$ $\l$-deformation explained above. Pursuing in the same way, the Lax  pair can be represented, using the untwisted basis \eqn{eq:generatorsun}, as
\begin{equation}
{\cal L}_\pm (z) =   \alpha(z^{\pm 1}) ( \tilde{A}_\pm^1 \tilde{t}_1 + \tilde{A}_\pm^2 \tilde{t}_2 ) + \beta(z^{\pm 1}) \tilde{A}_\pm^3 \tilde{t}_3 +\gamma(\zeta) \tilde{A}_\pm^4 \tilde{t}_4\ ,
\end{equation}
with the equations of motion \eqref{eomAinitial} and the constraints \eqn{eq:GaugeConstraints} evaluated  for the matrix \eqn{eq:LambdaAni}  (enhanced with a fourth entry equal  to unity),   and with the coefficients given by
\begin{equation}
\begin{split}
&\alpha(z) =   \sqrt{ \frac{(1+\lambda)^2 z + q^2 \lambda (1+z)^2}{(1+q^2) \lambda (1+\lambda)^2 z^{2}}} \ ,
\\
& \beta (z)  =  \frac{(1+\lambda)^2 z + 2 q^2 \lambda (1+z)}{(1+\lambda) (1+\lambda + 2 q^2 \lambda) z } \ ,
 \\ 
&\gamma(\zeta) =  \zeta\ ,  
\end{split}
\end{equation}
where $z,\zeta \in \mathbb{C}$ remain free. The additional redundancy in $\zeta$ can be attributed to the residual gauge direction, since we may simply choose the gauge $\tilde A_\pm^4=0$. This is in accordance with  the anisotropic $SU(2)$ formulation of \cite{Sfetsos:2014lla}, in which there is no residual gauge, and the equations of motion only exhibit a single spectral parameter.


\subsection{Renormalization group flow}

For symmetric $G/H$ spaces the $RG$-flow of the deformation parameter was found by different methods in \cite{Itsios:2014lca,Appadu:2015nfa} and the same result holds for non-trivial automorphisms  as well \cite{Driezen:2019ykp}.
For non-symmetric spaces the $RG$-flow is more involved. The interested reader may consult 
several such cases in subsection 3.1 of \cite{Sagkrioti:2018rwg}. 
One issue that has not been definitively settled is whether an integrable model is a consistent truncation
and stays integrable under the $RG$-flow.
An additional complication here is that the group $G = SU(2)\times U(1)$ is non-semisimple and the general considerations of \cite{Sagkrioti:2018rwg}, and earlier work, can not be applied. Hence, we have to investigate the $RG$-flow properties of our examples on their own and find the corresponding running
for the parameters $\l$ and $q$ using the gravitational method and the explicit deformed geometry. 
The canonical equivalence between the axial and vector deformed theories allows one to choose either one in order to derive the one-loop $RG$-flow equations. We choose to work with
the vector model since in that case the background antisymmetric tensor is zero and the
computational task is seemingly easier.

As argued, we will calculate the $\beta$-functions of our model using the gravitational method.
For a general $\sigma$-model, the one-loop $\beta$-functions in the absence of torsion are given by \cite{Ecker:1972bm,Honerkamp:1971sh,Friedan:1980jf}
\begin{equation}
\frac{d G_{\mu\nu}}{{\color{black} d\log\mu^2}} =  R_{\mu\nu} + \nabla_{\mu} \xi_{\nu} +
\nabla_{\n} \xi_{\m} \ ,
\end{equation}
where $R_{\mu\nu}$ is the target space Ricci tensor, $\nabla$ is the usual Christoffel connection and $\xi^\m$ corresponds to possible reparametrizations along the flow needed for consistency.
In our case we use the vector geometry \eqn{eq:vectorLambda} and \eqn{eq:vectorLambda1} and   we make the following ansatz  $\xi_\mu (\omega , \phi , x) = \nabla_\mu \Phi + a_\mu (\omega , \phi) x$. By exploiting the integrability conditions of the system we find that we must take
\begin{gather}
\xi^x = - \frac{2\lambda\, x}{k(1-\lambda^2)}\ , \quad
 \xi^{\phi_1} = \frac{2\lambda\ \sin 2\phi_1 }{k(1-\lambda^2)} \ ,  \quad
  \xi^\omega = - \frac{ 1+\lambda^2 - 2\lambda \cos 2\phi_1   }{k(1-\lambda^2)}\,
  \cot\omega\ ,
\end{gather}
such that the renormalisation of parameters  does not introduce additional couplings at one-loop.
Note that after raising the index on $\xi_\mu$ with the vector metric, the components of the diffeomorphism have no dependence of the parameter $q$.
Then, it turns out that the $\beta$-function equations are given by
\be
\begin{split}
 \label{eq:RGFlow}
\frac{d\lambda}{dt} &= - \frac{ \lambda}{k (1+q^2)} - \frac{2 q^2 \lambda^2}{k(1+q^2) (1+\lambda)^2}\ ,
\\
\frac{dq}{dt} &= -  \frac{ q \lambda}{k (1-\lambda^2)}\ ,
\end{split}
\ee
where we defined the RG time $t = \log\mu^2$.
The level $k$ does not run which is consistent with the fact that it is an integer.
Let us point out that this result is to leading order in $1/k$ (one-loop) but at finite values of $\lambda$ and $q$. As a consistency check,  the above result is invariant under the discrete symmetries \eqref{eq:LambdaDisSymm} and \eqn{qx}
and we obtain the right behaviours in the limiting cases. For $\lambda \rightarrow 0$ and $q$ generic there is a line of fixed points in the UV corresponding to the gauged WZW models on $(SU(2)\times U(1))/U(1)_q$. For $q\rightarrow \infty$
the running of the deformation parameter $\lambda$ coincides with the result for the $\lambda$-deformed $SU(2)$ group manifold. For $q \rightarrow 0$, the running 
coincides with the results   for the $\lambda$-deformed $SU(2)/U(1)$ symmetric space \cite{Itsios:2014lca, Appadu:2015nfa}. In general, 
for small $\lambda$ the operator driving the deformation away from the conformal point is the parafermion bilinear of the theory.
For symmetric spaces the anomalous dimension of this operator suffices to determine completely the flow of the parameter
$\l$, i.e.~the linear in $\l$ term in \eqn{eq:RGFlow}. The additional term in that equation is presumably due to the
fact that  the space is non-symmetric. Interestingly, the $\beta$-function for $\l$ is in a sense a linear combination of the group and coset cases, weighted with $q$, and thus the running of $\l$ at one-loop receives both a semi-classical (from the coset action) and truly quantum (from the group) contribution. This is similar to examples provided in \cite{Sagkrioti:2018rwg}. 
In addition, as expected from the equivalence to the biaxial $SU(2)$ $\l$-deformed model demonstrated above, we have checked the equivalence of the corresponding $\beta$-functions. In particular, eq. (6.2) of \cite{Sfetsos:2014jfa} with $\l_1=\l_2=\l$ and $\l_3$ as in 
\eqn{eq:LambdaAni} gives rise to the system \eqn{eq:RGFlow} above.

\begin{figure}[H] 
\label{betaplot4}
\begin{center}
\vskip 0 cm
\includegraphics[height= 8 cm,angle=0]{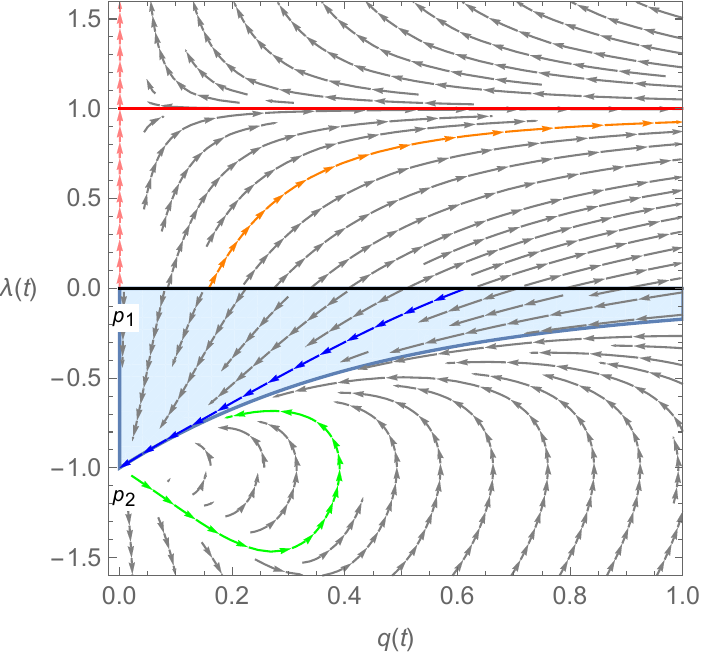}
\end{center}
\vskip -.7 cm
\captionsetup{width=.85\linewidth}
\caption{
RG evolution of $\{q, \lambda\}$ as given in \eqref{eq:RGFlow}. We have redefined the RG time $t$ to absorb the level $k$ which at one-loop effects only the rate of the flow. The black line for $\lambda = 0$ is the line of UV fixed points of gauged WZW models for generic $q$. The black dot $p_1$ represents the well-known $SU(2)/U(1)$ CFT, while $p_2$ is the point where the geometrical description breaks down. The red line for $\lambda=1$ defines a strong coupling regime corresponding to the non-Abelian T-dual of the $\sigma$-model (for the vector case) or its Abelian T-dual (for the axial case). The RG trajectories labeled as pink, orange,  blue and green have an RG invariant $\Theta = \infty$, $6359/4$, $7$ and $-26/4$, respectively.
\label{f:RG}}
\end{figure}

 In the remainder of this section we discuss the RG phase portrait of the system \eqref{eq:RGFlow} presented in fig.~\eqref{f:RG}  in which the arrows point towards the IR. The system has the RG invariant
 \begin{equation}
 \label{RGinv}
 \Theta =  \frac{1+ 2(1+2q^2) \lambda + \lambda^2 }{q^4 (1-\lambda)^2}\ ,
 \end{equation}
 which labels the RG trajectories and moreover is invariant under the discrete symmetries \eqref{eq:LambdaDisSymm} and \eqn{qx}.
 We relax the requirement that $\lambda\in [0,1]$, which is the physical region by  construction, by extending this interval to include both negative and arbitrarily large values.
In  the UV, we reach the  line of fixed points $\lambda = 0$ for constant parameter $q$ corresponding to the gauged WZW
CFTs. The line with $\l=1$ is the strong coupling region where the non-Abelian T-dual limit was taken.
This zoom-in limit where  the level $k$ tends to
 infinity, i.e.~as in \eqn{nonabellimit} and in \eqn{nonabellimit2}, is still valid when taken in the
 $\beta$-function equations \eqn{eq:RGFlow} and gives
 \be
{d\k^2\ov dt}= {2+q^2\ov 2(1+q^2)}\  ,\qq {dq\ov dt} = -{q\ov 2\k^2}\ .
\ee
The $RG$-flow invariant of this system is given by
\be
 \label{RGinvnab}
\th = {1+q^2\ov q^4 \k^4} \ ,
\ee
which follows by taking the zoom-in limit in the $RG$-flow invariant \eqn{RGinv}.
More generally, for $\l>0$ the conserved quantity $\Th$ is always positive. For negative $\l$ it is still positive in the blue
 shaded region and
negative below it.  For $\Th<0$ we see cyclic flows starting and ending at the point  $p_2$ with  $q=0$ and
$\lambda \rightarrow -1$. Although this resembles a fixed point,  for both the vector and the axial cases the $\sigma$-model description breaks down as the
determinant of the metrics vanishes identically. The occurrence of cyclic flows is consistent with \cite{Appadu:2017bnv}.

\section{Conclusions}
\label{conclusions}

In this work we provide an integrable deformation of the Euclidean three-dimensional black string using asymmetric $\lambda$-deformations on $SU(2)\times U(1)/U(1)_q$. We show that these models  continue to be canonically related by an axial-vector T-duality despite the fact that the deformations are not marginal.  Furthermore, we show that they have an equivalent description in terms of an anisotropic $SU(2)$ deformation, a feature
that one would not anticipate without the knowledge of the residual gauge symmetry. In the axial gauge formulation--which is the one that is easily transformed to a deformation of the Lorentzian black string--we proof classical integrability and one-loop renormalisability although in this case the space is non-symmetric and the Lie group is non-semisimple. This adds another example to  the apparent relation between integrability and renormalisability for two-dimensional sigma models, of which an understanding is emerging from e.g.~\cite{Hoare:2019ark,Hoare:2019mcc}  and references therein,  although here based on a non-symmetric space.

A peculiar feature of the gauge formulation is that the Lax pair admits a second spectral parameter. This clearly asks  for a more thorough integrability analysis, in particular  in terms of a twist function. It would be interesting to see whether this second spectral parameter could play a useful r\^ole in deriving the underlying symmetry properties of the deformed black string and possible extensions to deformed black brane backgrounds. In that respect for the anisotropic XXZ formulation, which has no residual gauge symmetry and as pointed out no second spectral parameter, the quantum group symmetries and the twist function were derived and studied in \cite{Appadu:2017bnv}.

A compelling future direction is to go to Lorentzian signature and study the \textit{blackness} of the deformed black string. In particular it would be interesting to analyse the gravitational implications, such as the near-horizon and asymptotic  geometry, of the deformation. At the undeformed CFT point $(\lambda = 0)$ the axial background \eqref{limitaxial} can be analytically continued to Lorentzian signature by transforming the parameters as $k \rightarrow - k$ and $ q^2 \rightarrow - (1+\rho^2)$ and the coordinate variables as  $\omega \rightarrow i R , \ \phi_1 \rightarrow i X_0/\sqrt{1+\rho^2}$, and $\phi_2 \rightarrow i X_1/\rho$. In this case one finds  the semi-classical Horne-Horowitz black string solution in terms of the variables of  \cite{Sfetsos:1992yi}. The original  background  \cite{Horne:1991gn} is obtained after the transformation \cite{Sfetsos:1992yi} $\cosh^2 R = \frac{r_+ - r}{r_+ - r_-}$, $X_0 \rightarrow  t/\sqrt{k}$ and $X_1 \rightarrow  x/\sqrt{k}$, and defining
\begin{equation} \label{eq:TransfHH}
\begin{gathered}
r_+ = M \equiv \sqrt{\frac{2}{k}} (1+\rho^2) e^a, \qquad r_- = \frac{Q^2}{M} \equiv  \sqrt{\frac{2}{k}}\rho^2 e^a , \qquad \rho^2 = \frac{r_-}{r_+ - r_-}\ ,
\end{gathered}
\end{equation}
after which one finally gets the Horne-Horowitz black string metric
\begin{equation}
d s^2 = - \left( 1- \frac{r_+}{r} \right) dt^2 + \left(1- \frac{r_-}{r} \right) dx^2 + \frac{k}{4 r^2} \left( 1- \frac{r_+}{r} \right)^{-1} \left( 1- \frac{r_-}{r} \right) ^{-1} dr^2\ . 
\end{equation}
 Note that $r$ is extended to a global coordinate taking all positive values.
At $r = 0$ there is the line of curvature singularities and $r = r_\pm$ are the event and inner horizons, respectively. The constant $a$ represents the constant part of the dilaton, whereas $M$  the mass and $Q$ the charge of the black string. The fact that this geometry has two isometries in $t$ and $x$ implies  that this metric represents  a charged black  string which is straight and static. Interestingly, performing the same set of analytic continuations and transformations on the deformed background results in an 
RG phase portrait  where the undeformed gauged WZWs are still located in the UV. 
A further important project is to embed the deformed background in type-II supergravity using a suitable ansatz for the RR fields. This is necessary to derive the physical parameters such as the deformed charge and mass of the black string using the low energy supergravity action.

Finally, it would be interesting to study the usage of non-trivial outer automorphisms in other coset spaces  admitted by the Dynkin diagram of $G$. For instance,
interesting cases include the CFT and  the static configuration of NS5-branes on a circle which arise, respectively, from the asymmetric \cite{Nappi:1992kv} or null \cite{Sfetsos:1998xd} gauging of $H = U(1)\times U(1)$ in $G = SL(2,R)\times SU(2)$. A more direct study is to compare the effect of the  outer automorphism on the $\beta$-functions of the deformed non-symmetric Einstein space $SU(3)/U(1)^2$  found in \cite{Sagkrioti:2018rwg} (this coset has importance in ten-dimensional string compactifications). 

\subsection*{Acknowledgements}

We are grateful to K.~Siampos for collaboration on the relation of our models to those in \cite{Sfetsos:2014lla}. 
We also acknowledge discussions with D.C.~Thompson and G.~Georgiou.  The research work of SD is supported by the grant of ``la Caixa'' Foundation (ID 100010434) with code LCF/BQ/PI19/11690019, 
by AEI-Spain (FPA2017-84436-P and Unidad de Excelencia Mar\'\i a de Maetzu MDM-2016-0692), 
by Xunta de Galicia (Centro singular de investigaci\'on de Galicia accreditation 2019-2022), and   by European Union ERDF. Additionally, SD acknowledges
   support by  COST (European Cooperation in Science and Technology) through the Action MP1405 QSPACE for a Short Term Scientific Mission to the National and Kapodistrian University of Athens, as well as the welcoming hospitality  of the same University.  The research work of KS was supported by the Hellenic Foundation for
Research and Innovation (H.F.R.I.) under the ``First Call for H.F.R.I. Research Projects to support Faculty members and Researchers and
the procurement of high-cost research equipment grant'' (MIS 1857, Project Number: 16519).


\providecommand{\href}[2]{#2}\begingroup\raggedright\endgroup

\end{document}